# Turning CVEs into Educational Labs: Insights and Challenges

Trueye Tafese

Canisius University, Buffalo New York

Master of Science in Cybersecurity

December 16 ,2024

**Abstract:** *This research focuses on transforming CVEs to hands-on educational lab for cybersecurity training. The study shows the practical application of CVEs by developing containerized lab environments- Docker to simulate real-world vulnerabilities like SQL Injection, arbitrary code execution, and improper SSL certificate validation. These labs has structured tutorials, pre- and post-surveys to evaluate learning outcomes, and remediation steps. Key challenges included interpreting limited CVE data, resolving technical complexities in lab design, and ensuring accessibility for diverse learners. Despite these difficulties, the findings highlight the use of educational benefits of vulnerability analysis, bridging theoretical concepts with hands-on experience. The results indicate that students improved comprehension of cybersecurity principles, threat mitigation techniques, and secure coding practices. This innovative approach provides a scalable and reproducible model for integrating CVEs into cybersecurity education, fostering a deeper understanding of real-world security challenges in a controlled and safe environment.*

# Table of Contents

# Introduction

Common Vulnerabilities and Exposures (CVE) is a publicly accessible database that identifies and catalogs known security vulnerabilities in software and hardware. Each vulnerability is assigned a unique ID, making it easier for organizations to share information, prioritize fixes, and protect their systems [1].

In 1999, the CVE system was established to make it easier to identify security problems. This system is managed by MITRE Corporation and supported by the U.S. government. It provides a common way to share and compare information about security issues. Before CVE, different companies had their own databases with different ways of identifying problems, making it hard to compare information. CVE fixed this by providing a standard system that everyone can use. The main purpose of the CVE system is to make it easy for groups to share information about holes and risks in security and work together to fix these problems [1].

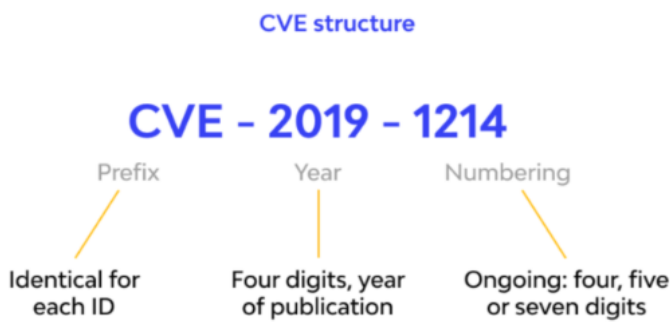

Figure 1: CVE Structure

Often vulnerabilities can take many different forms, Sensitive Data Exposure, SQL Injection, Code Execution, Denial of Service, Cross-Site Scripting (XSS), External Entity Injection, and Overflows are some frequent types. Due to their frequent occurrence in the field of cybersecurity, these categories are essential to comprehend and handle [2].

CVEs provide a standardized identification system for known cybersecurity vulnerabilities. By analyzing CVEs, learners can gain practical insights into how vulnerabilities are discovered, exploited, and patched, while developing essential skills in threat assessment, mitigation strategies, and secure coding practices.

The primary objectives of this research are to transform CVEs into practical, hands-on laboratory exercises for cybersecurity education and evaluate their effectiveness through student surveys. This involves developing a systematic approach for converting real-world vulnerabilities into interactive learning modules, creating comprehensive lab materials, and assessing student learning outcomes through quantitative survey methods to measure knowledge retention, skill development, and overall educational impact.

The scope of this research encompasses the development of containerized cybersecurity labs using Docker, specifically focusing on three critical vulnerability types: SQL Injection, arbitrary code execution, and improper certificate validation. The lab environment includes vulnerable Python applications and comprehensive step-by-step tutorials that guide students through the complete vulnerability lifecycle - from understanding CVE descriptions to practical exploitation and remediation. The significance of this approach lies in bridging the gap between abstract CVE documentation and practical implementation, enabling students to gain hands-on experience in a controlled sandbox environment. This methodology is particularly valuable as it provides a safe, reproducible platform for students to explore real-world vulnerabilities while developing practical skills in vulnerability assessment, exploitation, and mitigation.

# Literature Review

## CVEs in Cybersecurity Education

This literature review synthesizes insights from key studies, highlighting their contributions to educational methodologies and practical understanding of CVEs.

**Cybersecurity Defenses: Exploration of CVE Types through Attack Descriptions** [8] study introduced VULDAT, a machine learning-based classification tool that links CVE data with attack techniques. Using the MPNET sentence transformer, VULDAT made classification accuracy, to help understand the role of CVEs in broader attack frameworks. This offers students a tangible connection between CVE identifiers and practical cybersecurity threats.

**The Generic Error SDP and Generic Error CVE** [9] paper provided a theoretical foundation for CVE schemes, focusing on generic error sets and their mathematical and algorithmic properties. While primarily centered on secure coding and cryptographic error detection, it offers insights into the complexities of CVE schemes. This helps students to understand secure systems design.

**The CPE-Identifier: Automated CPE Identification and CVE Summaries Annotation** [4] research developed a deep learning-based system for annotating Common Platform Enumeration (CPE) entities in CVE summaries. By achieving high precision and recall, this tool simplifies associating CVEs with specific software platforms. This makes it an essential resource for teaching vulnerability management in diverse technological ecosystems.

**Automated Labeling of Entities in CVE Vulnerability Descriptions with NLP by applying Named Entity Recognition (NER) techniques to CVE descriptions** [10]: this research improved the semantic structuring of CVE data, achieving an F1 score of 0.93. This automated approach simplifies the analysis of CVEs and makes it efficient for teaching vulnerability classification and mitigation.

**Automating the Generation of NVD CVE Fixes Using Generative AI** [3] paper proposed a system combining generative AI and vector databases to automate the generation of CVE fixes, drastically reducing manual effort in vulnerability management. This innovation demonstrates how artificial intelligence can aid in real-world cybersecurity challenges.

## Studies and challenges on Hands-on Labs and Learning Effectiveness

Recent studies highlight the impact of hands-on laboratory experiences on learning outcomes in technical fields. Vykopal et al. (2021) [5] explored scalable teaching methods for cybersecurity labs, showing that hands-on activities greatly enhance practical skills but face challenges in scaling due to preparation demands, class management, and feedback. Their solution involved virtual networks supporting both local and cloud deployment, providing flexibility and scalability.

The Third IASTED International Conference [6] underscored the value of virtual machine-based labs in computer security education. Benefits included quick configuration resets, improved maintainability, and accessibility for larger groups, effectively addressing traditional limitations while retaining educational value.

Research from 2013 to 2023 charts the evolution of virtual labs from supplementary tools to essential learning platforms, especially during the COVID-19 pandemic. These labs enhance accessibility, scalability, and safety while supporting inquiry-based learning. Studies emphasize the need for balancing scalability, effective assessments, timely feedback, and the integration of theory with practice. Hybrid models combining traditional and virtual labs emerge as a comprehensive approach to modern education.

Designing realistic and understandable labs for cybersecurity education presents significant technical and pedagogical challenges. Scalability is a key issue, requiring dynamic environments that emulate real-world systems while accommodating diverse class sizes and learning needs. Realism often demands robust infrastructure, including virtual networks and operating systems, but these can strain resources, particularly in large classes.

Labs must simulate complex scenarios, such as attack response or CVE analysis without overwhelming students. Tools like AI-driven CVE fixes and NLP-based labeling systems offer promise but introduce additional complexities.
Providing immediate, actionable feedback is also the main problem. This shows the need for systems that track progress and enhance engagement.

In conclusion, these literatures highlight progress in cybersecurity education and CVE research but reveals a gap in integrating these efforts into accessible platforms like GitHub. While virtual labs and cloud systems dominate, more practical and inclusive implementations are needed. This paper is a great start for transforming CVE in a scalable learning environment.

# Methodology

This section will discuss the description of the selected CVEs, steps to transform CVEs into labs. It covers the analysis of vulnerabilities, designing and implementing controlled lab environments, and assessing their impact through pre- and post-surveys. This section will also highlight tools and technologies used throughout the process.

## Description of CVEs

### Case Study 1: Arbitrary code execution CVE-2017-18342

CVE-2017-18342 has a security vulnerability in the PyYAML library, affecting versions before 5.1. This vulnerability stems from the unsafe loading of YAML files, which can lead to arbitrary code execution.

**Technical Details:**

- Affected Component: PyYAML library
- Affected Versions: All versions prior to 5.1
- Vulnerability Type: Arbitrary Code Execution
- CVSS Score: 9.8 (Critical)
- Attack Vector: Network
- Attack Complexity: Low

The vulnerability occurs during the parsing of YAML files using the yaml.load() function. When processing YAML input, the function can instantiate arbitrary Python objects, potentially leading to code execution. This is particularly dangerous because:

- YAML files can contain serialized Python objects

- During deserialization, these objects are automatically constructed
- Malicious actors can craft YAML content that executes arbitrary code during the deserialization process

**Attack Scenario:**

- An attacker creates a specially crafted YAML file containing malicious Python code
- The application processes this file using yaml.load()
- During parsing, the malicious code is executed with the same privileges as the Python process

**Impact:**

- Remote Code Execution
- System Compromise
- Data Breach
- Service Disruption

**Mitigation:**

The primary mitigation strategy is to use yaml.safe_load() instead of yaml.load(). The safe_load() function restricts the types of objects that can be deserialized, preventing arbitrary code execution.

This vulnerability highlights the importance of secure deserialization practices and the need for careful handling of user-supplied input in applications that process YAML data.

## Case Study 2: Improper certificate validation CVE-2019-11324

CVE-2019-11324 has a security vulnerability in the urllib3 library, a python client that handles HTTP client. This vulnerability is related to improper certificate validation in the SSL/TLS verification process.

**Technical Details:**

- Affected Component: urllib3 library

- Affected Versions: Versions prior to 1.24.2
- Vulnerability Type: Improper Certificate Validation
- CVSS Score: 7.5 (High)
- Attack Vector: Network

The vulnerability arises from a flaw in the way urllib3 handles SSL certificate verification. Specifically, when urllib3 is configured to use custom SSL certificates:

- The library may accept certificates without properly validating the hostname.
- This occurs even when the SSL verification flag is explicitly set to True.
- The vulnerability affects systems using urllib3 with custom SSL configurations

**Attack Scenario:**

- An attacker sets up a malicious server with an invalid SSL certificate
- The vulnerable application attempts to make a secure connection
- Despite SSL verification being enabled, the connection is established without proper hostname validation
- This allows potential Man-in-the-Middle (MITM) attacks

**Impact:**

- Potential exposure to MITM attacks
- Compromise of sensitive data in transit
- Breach of secure communication channels
- False sense of security despite SSL implementation

**Mitigation:**

The primary mitigation strategies include:

- Upgrading to urllib3 version 1.24.2 or later
- Implementing proper certificate validation checks
- Using certificate pinning where appropriate

- Regular security audits of SSL/TLS configurations
- Implementing additional layers of security for sensitive communications

This vulnerability emphasizes the critical importance of proper SSL certificate validation and the potential risks of relying solely on default security configurations in networking libraries.

## Case study 3: SQL Injection

For the SQL demonstration, I focused on two CVEs: CVE-2018-20061 and CVE-2024-5725. The first, CVE-2018-20061, is a SQL injection vulnerability in ERPNext (versions 10.x and 11.x up to 11.0.3-beta.29). Exploiting this flaw requires only a basic user login without special privileges. The vulnerability lies in the /api/resource/Item?fields= endpoint, where attackers can use malicious JavaScript function calls to execute server-side Python functions with specially crafted arguments. This allows attackers to construct SQL queries that retrieve any column from any table in the database [13].

The second, CVE-2024-5725, is a SQL injection vulnerability in Centreon that enables authenticated attackers to execute arbitrary code. It stems from improper validation in the initCurveList function, allowing attackers to inject malicious SQL strings. If exploited, it permits code execution with Apache user privileges, compromising the system. This vulnerability, tracked as ZDI-CAN-22683, shows how inadequate input validation can lead to severe breaches despite authentication controls [14].

In my research, I attempted to replicate these vulnerabilities but faced challenges due to the unavailability of the affected versions in public repositories. ERPNext version 11.0.3-beta.29 is no longer accessible on Docker Hub, and Centreon's vulnerable version has been removed for security reasons. This limitation highlights a common obstacle in security research especially for SQL Injection, where outdated versions are intentionally removed to prevent misuse. Unfortunately, this practice also obstructs legitimate research and education, making it impossible to recreate these vulnerabilities for demonstration in controlled environments.

## Steps to transform CVEs into labs

The goal of this project is to create a lab environment accessible to everyone by using CVEs as a foundational resource for understanding cybersecurity and security vulnerabilities. The process involves the following steps:

1. Research Focus Area: I choose the types of vulnerability to implement, the environment to deploy and the programming language to use for automation. For this paper,I used python and Docker containers as a sandbox environment.

2. Initial Analysis: I evaluated each CVE for its educational value by asking questions such as "Does it cover significant cybersecurity topics?", "Is the implementation too complex for students?", and "What can students learn and apply from this?".

3. Lab Design: After choosing a CVE based on the analysis, I translated it into a Dockerized environment. By parsing the CVE and associated exploit details, I created Docker Compose files that simulated the vulnerable setup. This step allowed students to interact with real-world scenarios in a controlled environment.

4. Exploit Scripts: I developed blueprint exploit scripts tailored to the vulnerabilities. These scripts enabled users to test and interact with the vulnerabilities in real-time, providing a hands-on learning experience.

5. Remediation Strategies: After running the exploits, I provided guidance on remediation techniques. This included steps such as patching, adjusting configurations, and implementing best practices to mitigate similar threats in production environments.

6. Scalability and Reusability: To ensure broader accessibility, I published the Docker-based lab setups on GitHub. This approach allows the labs to be reused and scaled for various applications, including classrooms, workshops, and enterprise training programs.

7. Step-by-Step Tutorial: Finally, I compiled all the processes into a step-by-step tutorial. This guide ensures that users can easily follow along, replicate the setups, and gain a comprehensive understanding of the concepts. By documenting the entire workflow, the tutorial enhances accessibility and serves as a valuable resource for learners and instructors alike.

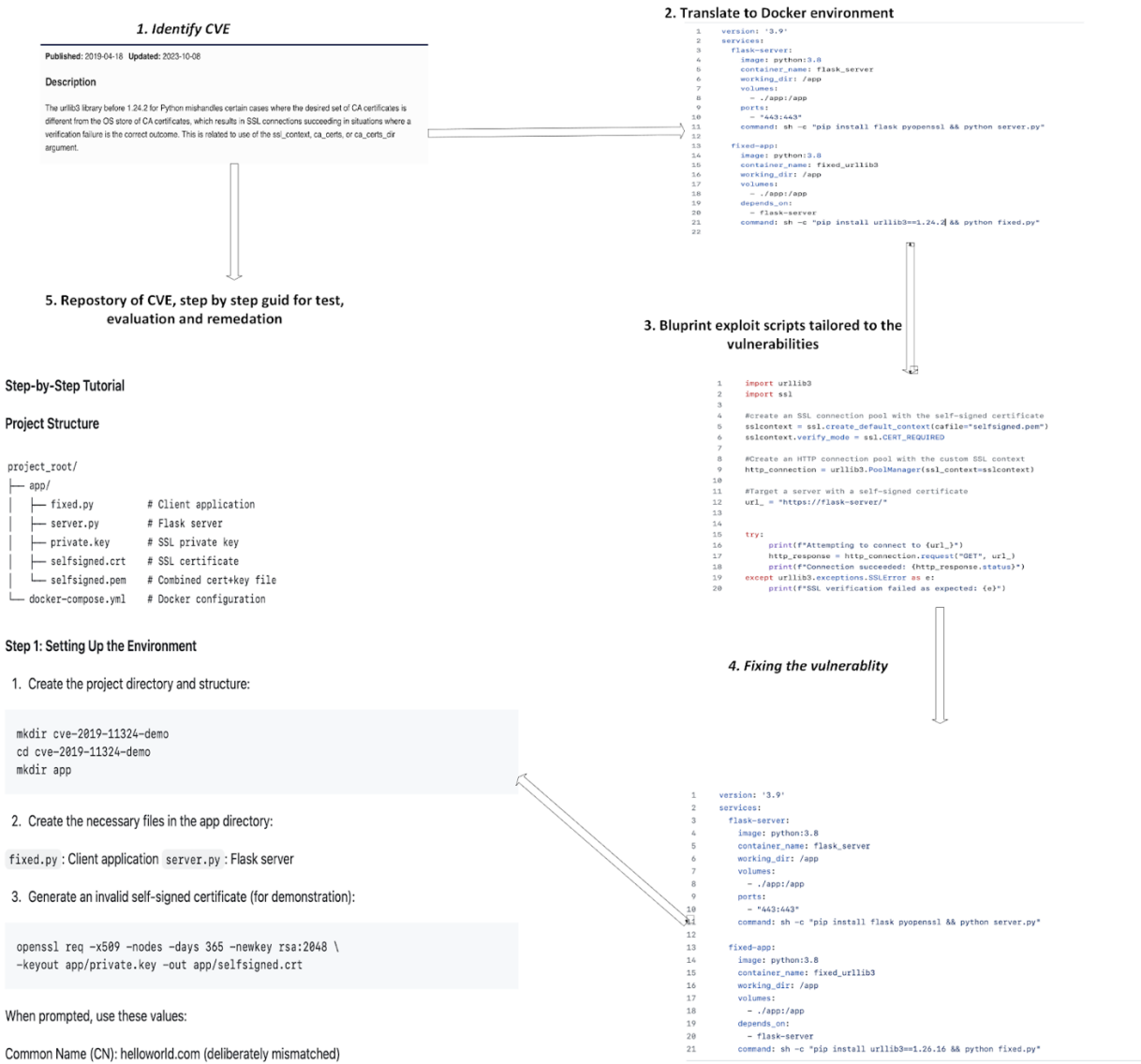

figure 2: *Develop an Effective, Realistic Cyber Learning Environment*

## Pre- and post-survey structure and purpose

The combined pre- and post-survey evaluates students' understanding of CVEs, vulnerability exploitation, and remediation. It assesses knowledge on technical implementation, code comprehension, security awareness, and practical application. The survey also gathers feedback for tutorial improvement and analyzes the tutorial's effectiveness for future use in undergraduate cybersecurity courses.

## Tools, programming language and technologies used.

- Docker: Used for containerization
- Visual Studio Code: A code editor to write and edit scripts and configurations.
- Unix OS: The operating system used for development and testing.
- GitHub: For version control, sharing code, and creating the README tutorial.
- Python: Used to automate exploitation and simulate attacks.
- YAML: To define the docker compose file for environment configuration.
- Google Forms: For creating surveys to gather feedback or data.
- CVE MITRE: To find CVE Common Vulnerabilities and Exposures
- VISio: To create figures

# Findings

## Lab Development Challenges:

The development of CVE-based laboratories shows some challenges. A key issue was the limited information in CVE entries, which does not offer brief descriptions and basic references. For example, the CVE-2017-18342 (PyYAML vulnerability) entry only mentioned arbitrary code execution without detailing the specific YAML parsing mechanisms that could be exploited. This required extensive research across multiple sources like PyYAML's documentation, related GitHub issues, security advisories, and the NIST National Vulnerability Database, to fully understand the attack vector.

The CVEs system's primary focus on unpatched software vulnerabilities overlooks other challenges. During the PyYAML and urllib3 implementations, I encountered related security issues, like configuration mistakes not covered by CVE entries. This broader scope of risk management, which CVEs alone cannot address, made the process of creating comprehensive laboratory environments more complex.

Another significant problem was the reliance on external sources for comprehensive analysis. In the process of developing the lab, I had to gather information from multiple sources to determine the affected versions and exploitation mechanisms. Similarly, creating the urllib3 or PyYAML lab required analyzing real-world exploitation scenarios and consulting various security advisories. This fragmentation of essential information delayed the development of accurate and educational laboratory environments, underscoring the need for a more integrated approach to vulnerability information management.

## Survey Results: Student Performance Pre- and Post-Labs

Pre-Lab Performance

Most students had limited prior exposure to cybersecurity vulnerabilities, reporting minimal familiarity with CVE-2017-18342 (YAML deserialization) and CVE-2019-11324 (SSL certificate validation). Initial challenges included grasping technical concepts (e.g., `yaml.safe_load()` and Docker configurations) and overcoming setup difficulties.

Post-Lab Improvements

Students reported significant gains in understanding critical concepts, such as safe deserialization, SSL/TLS security, and Docker-based testing. Confidence levels also improved, with many students feeling prepared to identify and mitigate vulnerabilities.

with many students feeling prepared to identify and mitigate vulnerabilities.

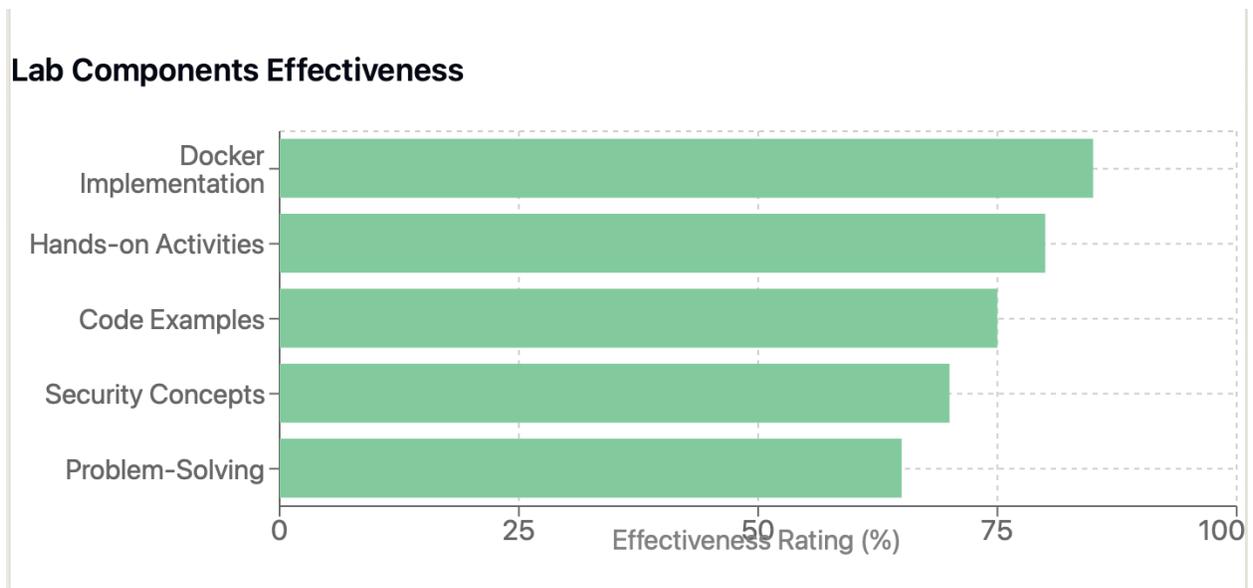

Figure 3: pre vs post Lab understanding

Student suggestions for Improvement

- Additional code examples and exercises
- Real-world case studies
- Video demonstrations
- Integration of tools like Wireshark for better visualization

Impact and Feedback

The labs effectively bridged theory and practice. Students reported increased confidence in auditing projects and adopting best practices, such as updating libraries and ensuring secure configurations. Hands-on activities were praised for demonstrating real-world applicability.

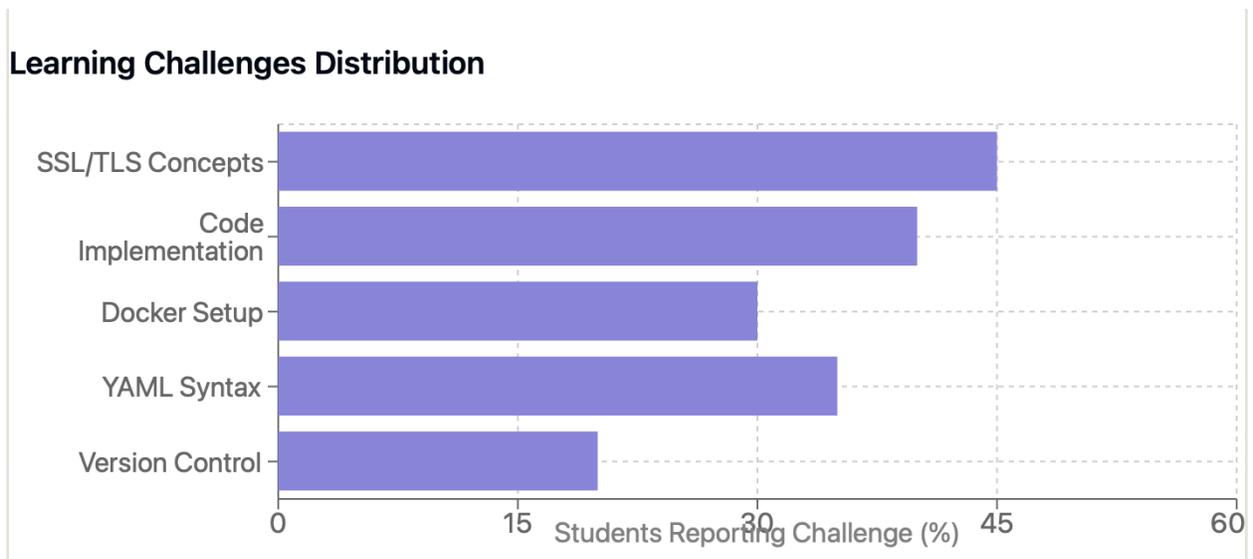

Figure 4: Learning challenge distribution

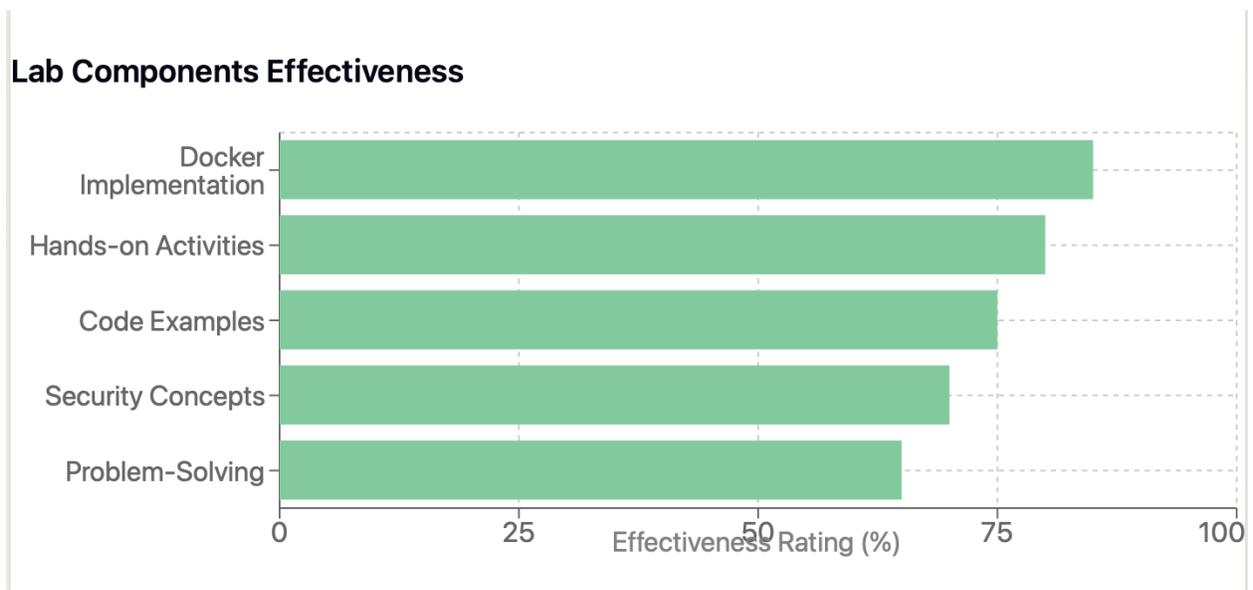

Figure 5: Lab component effectiveness

In conclusion, interactive lab tutorials significantly enhanced students' cybersecurity knowledge and practices. Addressing learning gaps and expanding resources can further amplify their impact, reinforcing the value of experiential learning in cybersecurity education.

The survey results indicate that the hands-on lab approach successfully achieved its educational objectives, with participants gaining practical experience and theoretical understanding of both vulnerabilities. The Docker-based implementation proved particularly effective in providing a safe, controlled environment for security education.

# Discussion

This research highlights how lab-based tutorials improve student understanding of cybersecurity vulnerabilities. By engaging with hands-on demonstrations, students become familiar with CVEs, bridging the gap between theoretical descriptions and practical application. This research also demonstrates the feasibility of transforming CVE descriptions into functional code demonstrations, which can serve as the foundation for creating educational environments accessible through GitHub.

These labs do not only introduce CVEs to students, but they also enhance coding skills and deepen understanding of Python libraries. For instance, students learn concepts such as certificate validation and secure deserialization through practical exercises. This approach equips students with a strong understanding of secure software engineering principles.

However, there are some challenges. A significant issue is the difficulty of replicating certain demonstrations due to the unavailability of Docker Hub images, as some are removed for security reasons. This highlights the need for curated and secure Docker images specifically designed for educational purposes. Another issue is the lack of detailed resources and documentation for certain CVEs hinders comprehensive learning.

Another thing that is observed in this paper is students grasp the mechanics of attacks and defenses. For some, abstract concepts remain difficult to understand without additional visualization. Incorporating user interfaces or visualization tools could provide clearer representations of how attacks work and their impact. For example, connecting Docker container into a graphical UI or relating examples to real-world scenarios could enhance comprehension.

Future implementations should focus on creating advanced and realistic environments. These environments could integrate Docker with user-friendly interfaces, offering students an intuitive way to explore vulnerabilities and defenses. Providing modular and reusable resources on platforms like GitHub could further increase accessibility and scalability.

# Conclusion

In this research, we explored critical vulnerabilities, namely CVE-2019-11324 and CVE-2018-20061, to demonstrate the importance of identifying and remediating security issues in modern software systems. Our findings highlight the significance of proper SSL certificate verification in the urllib3 library, and the risks posed by SQL injection vulnerabilities in ERPNext. Through simulation in controlled Docker Compose environments, we validated the exploitation techniques and proposed practical remediation steps, underscoring their relevance in cybersecurity research and training.

Future research could focus on expanding the scope to include more CVEs, especially those impacting large-scale enterprise systems, to enhance the depth of security assessments. Moreover, integrating automated tools for vulnerability scanning and remediation could improve both the efficiency and accuracy of these practices. By continuing to investigate and mitigate software vulnerabilities, researchers and practitioners can contribute significantly to building a safer digital landscape.

# Appendix

1. Github Repository for CVE-2019-11324 https://github.com/Tiruye-z/CVE2Lab/tree/main/CVE-2019-11324

2. Github Repository for CVE-2017-18342 https://github.com/Tiruye-z/CVE2Lab/tree/main/CVE-2017-18342
3. Survey Questionnaire for CVE-2019-11324 https://forms.gle/h9D1g8KA1zaWPstm7
4. Survey Questionnaire for CVE-2017-18342 https://forms.gle/4RBPkYHDfZkaGCFS8